# Generative Modeling and Decision Fusion for Unknown Event Detection and Classification Using Synchrophasor Data

Yi Hu, *Member,* and Zheyuan Cheng, *Member, IEEE*

*Abstract*—Reliable detection and classification of power system events are critical for maintaining grid stability and situational awareness. Existing approaches often depend on limited labeled datasets, which restricts their ability to generalize to rare or unseen disturbances. This paper proposes a novel framework that integrates generative modeling, sliding-window temporal processing, and decision fusion to achieve robust event detection and classification using synchrophasor data. A variational autoencoder–generative adversarial network (VAE–GAN) is employed to model normal operating conditions, where both reconstruction error and discriminator error are extracted as anomaly indicators. Two complementary decision strategies are developed: a threshold-based rule for computational efficiency and a convex hull–based method for robustness under complex error distributions. These features are organized into spatiotemporal detection and classification matrices through a sliding-window mechanism, and an identification and decision fusion stage integrates the outputs across PMUs. This design enables the framework to identify known events while systematically classifying previously unseen disturbances into a new category, addressing a key limitation of supervised classifiers. Experimental results demonstrate state-of-the-art accuracy, surpassing machine learning, deep learning, and envelope-based baselines. The ability to recognize unknown events further highlights the adaptability and practical value of the proposed approach for wide-area event analysis in modern power systems.

*Index Terms*—*Decision Fusion, Event Detection and Classification, Generative Artificial Intelligence, Phasor Measurement Units, Sliding Window, Variational Autoencoder-Generative Adversarial Network.*

## I. INTRODUCTION

WITH the growing penetration of renewable energy, increasing system complexity, and tighter operating margins, the need for real-time, accurate, and robust detection and classification of power system disturbances has become critical. Synchrophasor data, collected via Phasor Measurement Units (PMUs), offers high-resolution, time-synchronized measurements that are ideal for monitoring system dynamics and enabling early warning of abnormal events, which enables effective situational awareness (real-time event detection and classification), in order to avoid large-scale blackouts [1][2]. However, effectively analyzing such data at scale remains a significant challenge.

Existing approaches to PMU event detection and classification can be broadly categorized into signal processing-based and data-driven methods. Traditional signal processing techniques, such as wavelet transforms[3], short-time Fourier transforms[4], and principal component analysis (PCA)[5], detect events by examining frequency-domain components or low-rank subspace deviations in raw PMU data. Other methods employ correlation coefficient matrices [6] or singular value decomposition (SVD) [7] to capture spatial correlations and structural changes. Statistical feature extraction is used to classify events from PMU data stream[8]. A dynamic programming-based swinging door trending (DPSDT) method is proposed for high-precision PMU event detection [9]. While effective in some settings, these rule-based methods typically depend on manually defined features, thresholds, and fixed-time windows, which limit their adaptability and accuracy for detecting diverse event types.

To improve generalization, supervised learning models have been introduced, including decision trees[10], convolutional neural networks (CNNs) [11][12][13][14], and hybrid models such as autoencoders with LSTMs [15][16]. These approaches require both positive and negative labeled samples, which can be difficult to obtain in realistic grid environments, particularly for rare or emerging event types. Furthermore, many supervised models rely on extracted features rather than directly utilizing raw PMU signals, potentially losing critical temporal and spatial resolution. The recent use of GANs for PMU event detection [17] marks a shift toward unsupervised generative modeling, but conventional GANs face limitations in training stability and latent space interpretability.

A major challenge in power system event classification is the scarcity and imbalance of labeled event data. Many disturbances, such as rare faults or atypical oscillations, occur infrequently in practice, limiting the effectiveness of supervised learning approaches. Consequently, numerous studies on PMU-based event classification rely on only small datasets with limited labeled events. For example, [7] employs just 32 labeled events, [10] uses case studies with four PMUs, [18] trains an event classifier on only 57 labeled line events, and [11] utilizes several hundred labeled frequency events

This study was supported by the U.S. Department of Energy under Grant DE-EE0011375.
Yi Hu is with the Department of Electrical and Computer Engineering, Michigan Technological University, Houghton, MI, 49931, USA (e-mail: yhu6@mtu.edu). He was with Quanta Technology, Raleigh, NC, 27607, USA. This work was conducted during an internship at Quanta Technology.
Zheyuan Cheng is with Quanta Technology, Raleigh, NC, 27607, USA (e-mail: zcheng@quanta-technology.com).

from the FNET/GridEye system to develop a CNN-based detector. Such constraints often result in poor generalization of event classifiers when applied to unseen conditions. To mitigate data scarcity, recent works have explored the use of synthetic data generated from simulations or neural networks. In [19], simulated data with artificially injected noise are employed, while [20] leverages generative adversarial networks (GANs) to create synthetic event samples for training.

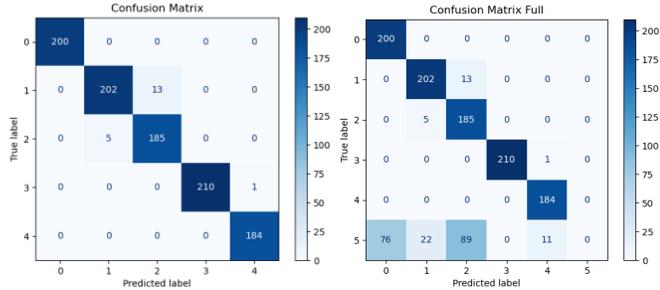

(a) Test on known categories.   (b) Test on unknown categories.
Fig. 1. Event classification accuracy degradation due to unseen event types.

Figure 1 illustrates the limitations of existing supervised classification methods. The confusion matrix on the left shows high accuracy (98.1%) when the model is trained and tested only on previously seen event types (Types 0–4). However, when a new event type (Type 5) is introduced at test time, as shown in the right-hand confusion matrix, the overall accuracy drops to 81.9%. This degradation occurs because the classifier, which relies on labeled training data, cannot generalize to previously unseen categories. Instead, samples from the new event type are incorrectly mapped to existing classes, as the misclassifications in the bottom row of the matrix. This example underscores two fundamental challenges: (1) many power system events are rare and difficult to capture in training datasets, and (2) supervised approaches inherently lack the ability to recognize novel or evolving disturbances. These challenges motivate the need for generative and unsupervised approaches that can learn from abundant normal-operation data and remain flexible in detecting and categorizing both known and unknown event types.

To address these limitations, this paper proposes a novel and generalizable event detection and classification framework based on Variational Autoencoder–Generative Adversarial Networks (VAE-GAN) using raw PMU data. The primary contributions of this paper are summarized as follows:

- *Application of VAE-GAN for power system events*: As far as we know, we are the first to apply a VAE-GAN model to synchrophasor data for both event detection and classification, leveraging its generative capability to capture the distribution of system behavior.
- *Unsupervised learning*: The proposed event detection framework requires only normal operating data for training, eliminating the need for large labeled event datasets, which are often scarce and imbalanced in practice.
- *Sliding-window and decision fusion for classification*: A novel sliding-window mechanism and decision fusion strategy are introduced to enhance event classification performance by exploiting spatiotemporal correlations and integrating detection and classification.
- *Handling unknown event types*: This framework is able to recognize and categorize previously unseen event types, surpassing conventional supervised classifiers that operate under closed-set assumptions.

The remainder of the paper is organized as follows: Section II provides a detailed introduction to the proposed method. Section III illustrates the simulation results. Finally, section VI concludes the paper.

## II. METHODOLOGY

The proposed methodology integrates unsupervised generative modeling with supervised learning to achieve robust event detection and classification in power systems using synchrophasor data. Unlike conventional supervised approaches that depend solely on labeled events and struggle with unseen types, the framework leverages generative AI to detect anomalies outside the distribution of normal data and categorize them as new event types. This design extends beyond closed-set assumptions, enabling more flexible and resilient event identification in real-world operations.

### A. Generative AI-Based Decision Fusion Framework

The proposed framework integrates unsupervised generative modeling, supervised classification with Multilayer Perceptron (MLP), sliding-window mechanism, and decision fusion to achieve robust event detection and classification using synchrophasor data. As illustrated in Figure 2, the framework begins with PMU data stream, which is processed by a VAE-GAN model to extract two error features. These error features are then passed through two parallel pathways. In the detection pathway, errors are evaluated by decision metrics to identify if there is an anomaly. In the classification pathway, the same error features are fed into an MLP that predicts event types for each segment. Then both pathways will be applied recursively by a sliding window mechanism, producing a binary detection matrix that represents spatiotemporal detection flags across PMUs and a non-binary classification matrix that indicates the event types.

Both matrices are subsequently processed in parallel through an identification stage: the binary detection matrix highlights whether each PMU segment deviates from normal operation, while the classification matrix assigns potential event categories to those same segments.

The outputs of these pathways are finally combined in the decision fusion module, where detection and classification results are jointly analyzed to determine the final event label.

Overall, this unified framework enables continuous real-time monitoring of wide area synchrophasor networks, with the capability to capture both short-lived transients and slowly evolving anomalies, while also generalizing to previously unseen event categories. The following sections introduce each component of the framework in detail.

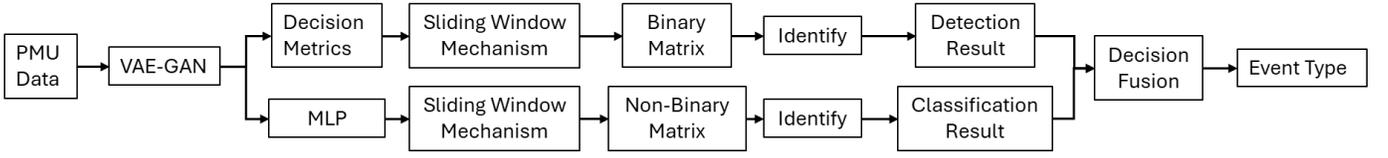

Fig. 2. Generative AI-based decision fusion framework.

## B. Signal Selection and Data Pre-processing

In the synchrophasor data, there are many signals that are available for event detection and classification. After extensive testing with field recorded synchrophasor data, the following signals provide the best detection and classification performance: Voltage magnitude at three phases $V_a$, $V_b$, and $V_c$, Positive sequence current magnitude $I$, Frequency $f$, Rate-Of-Change-Of-Frequency (ROCOF) $df$. The selected measurement from one PMU is represented as:

$$M_{PMU} = [V_a, V_b, V_c, I, f, df]$$
$$= \begin{bmatrix} v_a^1 & v_b^1 & v_c^1 & i^1 & f^1 & df^1 \\ v_a^2 & v_b^2 & v_c^2 & i^2 & f^2 & df^2 \\ \vdots & \vdots & \vdots & \vdots & \vdots & \vdots \\ v_a^N & v_b^N & v_c^N & i^N & f^N & df^N \end{bmatrix} \quad (1)$$

Then, the values for the $i^{th}$ time point are normalized by

$$v_{pu}^i = \frac{v^i}{v_{rate}}$$
$$v_{norm}^i = v_{pu}^i - \frac{1}{N}\sum_{i=1}^{N} v_{pu}^i$$
$$i_{norm}^i = i_m^i - \frac{1}{N}\sum_{i=1}^{N} i_m^i \quad (2)$$
$$f_{norm}^i = f^i - \frac{1}{N}\sum_{i=1}^{N} f^i$$

where $v^i$ is the phase voltage magnitude, $v_{rate}$ is the rated voltage at the measurement point, $i_m^i$ is the positive sequence current magnitude, and $f^i$ is the frequency. In this way, the normalized signals values are distributed around zero no matter the original voltage or current levels.

## C. Event Detection via VAE-GAN

This study utilizes a Generative Artificial Intelligence–based approach to detect power system events by modeling normal system behavior with a Variational Autoencoder–Generative Adversarial Network (VAE-GAN).

### 1) Model Configuration

As illustrated in Figure 3, the system consists of three major components: Encoder, Decoder/Generator, and Discriminator.

The encoder network receives normalized PMU measurements at a 5-second length $M$. It outputs two vectors: mean vector $\mu$, and standard deviation vector $\sigma$. A latent vector $z$ is then sampled using the reparameterization trick:

$$z = \mu + \sigma \cdot e, \quad where \ e \sim \mathcal{N}(0, 1) \quad (3)$$

The Decoder (Generator) reconstructs a signal $\widehat{M}_i = G(z_i)$ from the latent representation. This reconstruction is compared to the original input to detect anomalies. The Discriminator network receives both the original PMU signal and the reconstructed signal. It is trained to distinguish the difference between real and reconstructed samples, thereby enhancing the generator's realism.

The VAE-GAN model will take 5-second single PMU data $M$ as input and output two errors:

$$[e_{recon}, e_D] = VAEGAN(M) \quad (4)$$

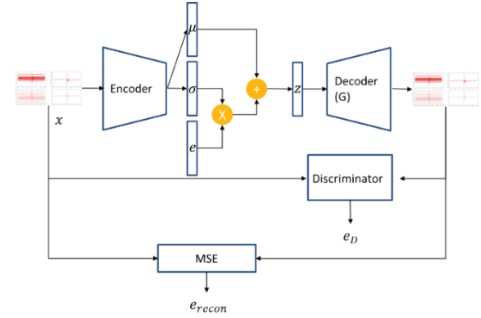

Fig. 3. VAE-GAN model architecture.

Under normal operating conditions, the model learns to accurately reconstruct time-series data from phasor measurement units (PMUs), resulting in consistently low reconstruction and discrimination errors. However, when an abnormal event occurs, the input pattern deviates from the learned distribution of normal behavior. As a result, the model's reconstruction becomes less accurate, and the discriminator identifies the difference more sharply, leading to a significant increase in error metrics. These deviations are used to flag and detect power system events.

### 2) Loss Function Designation for VAE-GAN

The total loss for training the VAE-GAN model is:

$$\mathcal{L}_{VAE-GAN} = \mathcal{L}_{recon} + \lambda_1 \mathcal{L}_{KL} + \lambda_2 \mathcal{L}_{adv} \quad (5)$$

where $\lambda_1$ and $\lambda_2$ are scaling coefficients. Reconstruction Loss that measures the difference between original data $M$ and reconstructed data $\widehat{M}_i$ using a hybrid of MSE and a max-penalty term:

$$\mathcal{L}_{recon} = \frac{1}{N}\sum_{i=1}^{N} \|M_i - \widehat{M}_i\|^2 + \max\left(\|M_i - \widehat{M}_i\|\right) \quad (6)$$

where $\widehat{M}_i = G(z_i)$. The MSE component provides a global measure of reconstruction fidelity across all samples, encouraging the model to minimize average reconstruction errors. However, in event detection tasks, rare but significant deviations are often masked when only the mean error is considered. To address this, we introduce the max-penalty term, which explicitly penalizes the largest reconstruction discrepancy within each window. This addition forces the model to pay attention to extreme deviations that may correspond to short-duration or localized disturbances, rather than averaging them out.

By incorporating this max-penalty item, the model becomes more sensitive to subtle but critical anomalies in synchrophasor data, such as sudden transient or rapid signal changes that might otherwise be overlooked. This hybrid loss design enhances the separation between normal and abnormal data in the error feature space, improving detection robustness. Importantly, the integration of the max-penalty into the reconstruction objective represents one of the key methodological contributions of this study, enabling the proposed framework to better capture the full spectrum of abnormal behaviors in power system event detection.

KL Divergence Loss regularizes the latent space:

$$\mathcal{L}_{KL} = -\frac{1}{2}\sum(1 + \log(\boldsymbol{\sigma}^2) - \boldsymbol{\mu}^2 - \boldsymbol{\sigma}^2) \quad (7)$$

Adversarial Loss part encourages realism in the reconstructed output. It is presented as:

$$\mathcal{L}_{adv} = -\mathbb{E}[\log D(G(\mathbf{z}))] \quad (8)$$

$\mathcal{L}_{VAE-GAN}$ is used for the training of Encoder and Decoder (Generator) part of the VAE-GAN model. The discriminator network $D$ in the VAE-GAN architecture is trained using:

$$\mathcal{L}_D = -\mathbb{E}[\log D(\boldsymbol{M})] - \mathbb{E}[\log(1 - D(G(\mathbf{z})))] \quad (9)$$

### 3) Decision Metrics

After extracting the reconstruction error $e_{recon}$ and discriminator output error $e_D$ from the VAE–GAN model, these two features are mapped to a 2D error feature space to determine whether the measured data corresponds to normal operation condition or an event condition. We develop two complementary decision-making strategies.

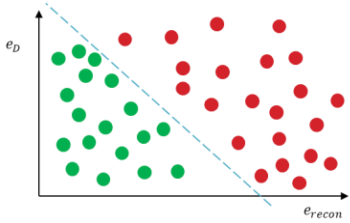

Fig. 4. 2D error feature space illustration.

**Method 1: Threshold based decision.** As illustrated in Figure 4, the threshold-based decision method can be viewed as a linear separation in the 2D error feature space. The reconstruction error $e_{recon}$ and discriminator output error $e_D$ form a two-dimensional plane where normal samples (green) cluster tightly, while abnormal samples (red) deviate from this region. By applying a weighted threshold, a linear boundary (dashed line) is established to distinguish between normal and event samples. An event is detected if the metrics meet the following criteria:

$$Event = \begin{cases} 1, & \lambda_{recon} \cdot e_{recon} + \lambda_D \cdot e_D > \eta_1 \\ 0, & otherwise \end{cases} \quad (10)$$

where $\eta_1$ is the threshold for event detection, $\lambda_{recon}$ and $\lambda_D$ are the weights for the two errors, respectively.

This visualization highlights the strength of the thresholding approach in its simplicity and interpretability but also underscores its limitation: when error distributions overlap or follow nonlinear patterns, a single linear threshold may be insufficient for robust separation. This motivates the introduction of the convex hull method, which adapts more flexibly to irregular data distributions.

**Method 2: Convex Hull based decision.** Instead of relying on a linear threshold, the convex hull method uses a geometric boundary to separate normal and abnormal samples in the error feature space. Specifically, a convex hull is constructed around the cluster of normal data points in the two-dimensional space defined by reconstruction error $e_{recon}$ and discriminator error $e_D$. During training, only non-event data is used to estimate this convex region. At testing stage, each new sample (or time window) is mapped into the same feature space. If the point lies inside the convex hull, it is considered consistent with normal operation; if it falls outside the hull, the sample is flagged as an event.

This approach has two advantages. First, it does not require explicit threshold tuning since the convex hull naturally captures the distribution of normal data. Second, it improves robustness when the distribution of errors is nonlinear or multi-modal, conditions under which a simple linear threshold may be insufficient. Mathematically, the decision rule can be expressed as:

$$Event = \begin{cases} 1, & x \notin \mathcal{C}(e_{recon}, e_D) \\ 0, & x \in \mathcal{C}(e_{recon}, e_D) \end{cases} \quad (11)$$

where $\mathcal{C}(e_{recon}, e_D)$ denotes the convex hull constructed from the training set of normal data.

Using the decision matrices, event detection can be achieved at the level of a single PMU within a relatively short time window (e.g., 5 seconds). However, system-wide event detection requires a broader temporal context. To capture events that evolve across multiple PMUs and over longer durations (e.g., 1 minute), a sliding window mechanism is necessary. This mechanism, introduced in the next section, provides the temporal aggregation needed to move from local, short-term detection to reliable wide-area event identification.

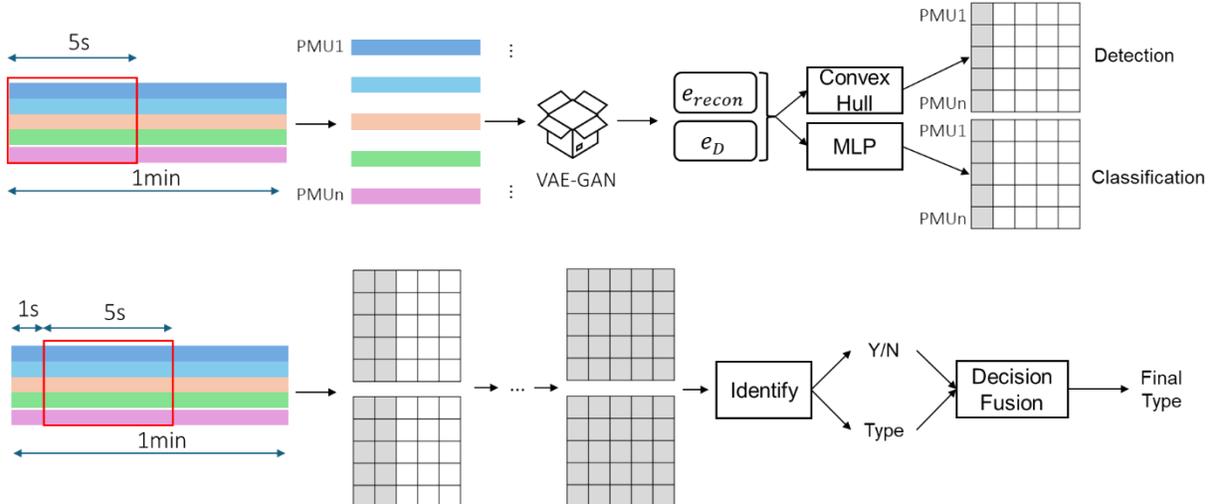

Fig. 5. The sliding window mechanism.

### D. Sliding Window-based Processing

A sliding window mechanism is incorporated to enable continuous and long-term monitoring of power system events using phasor measurement unit (PMU) data streams. This mechanism is essential to the temporal structure of the method, allowing it to detect both short-duration transient and slow-developing system events over extended periods. Most importantly, it enables the classification for event types that are not included in the training dataset.

As shown in Figure 5, A larger time window (e.g. 1 minute) of PMU data is segmented into smaller overlapping windows (e.g. 5-second segments with 1-second steps). Then, each short segment $M$ is independently processed by the VAE-GAN model to generate reconstruction error $e_{recon}$ and discriminator error $e_D$. These errors are then passed through two parallel decision modules: (1) Detection Module: The errors are evaluated using either the weighted-threshold method or the convex-hull method introduced in the previous section. This produces a binary abnormal matrix across PMUs and time segments, indicating whether each segment deviates from normal operation. (2) Classification Module: The same error features are also fed into a lightweight MLP, trained on the labeled event types. The MLP outputs a non-binary classification matrix, where each entry represents the predicted event type (or non-event) for the corresponding PMU and time segment.

This sliding window framework is not simply a pre-processing step, it is an integral part of the framework's ability to detect and classify events that unfold gradually across PMUs, capture temporal continuity in disturbances, and reduce false positives by incorporating spatiotemporal correlations. By structuring the input this way, the method ensured robust detection and classification of both localized and system-wide anomalies in real-time PMU streams.

An identification and decision fusion process will be applied to the binary detection matrix and non-binary classification matrix generated in this stage.

### E. Identify & Decision Fusion

The final stage of the framework combines the outputs of the detection and classification pathways to generate reliable system-level event decisions. This process involves two steps. First, an identification procedure is applied to the binary abnormality matrix and the non-binary classification matrix to assign event types across individual PMUs and time segments. Second, a decision fusion mechanism integrates these results to produce the final outcome, ensuring that both known and unseen events are consistently represented.

#### 1) Event Detection and Classification Identification

Figure 6 illustrates the identification process for event detection and classification based on the spatiotemporal matrices generated in the sliding windowing stage. For each PMU row in the matrix, the algorithm scans the sequence of outputs to identify the maximum number of consecutive occurrences of the same event type and records both the event label and its count. If this count exceeds a predefined threshold $\eta$, the row is assigned with that event type; otherwise, it is marked as non-event (Type = 0). This procedure is repeated for all rows until the last PMU has been processed.

The binary detection matrix differs from the non-binary classification matrix in its representation: the detection matrix contains only 0s and 1s, where 1 indicates abnormality, while the classification matrix contains values from 0 to 5, with 0 representing non-events and 1–5 corresponding to different event categories. Once the row-wise analysis is complete, the event types are aggregated across PMUs, and the most frequently occurring type is selected as the decision for that time slice.

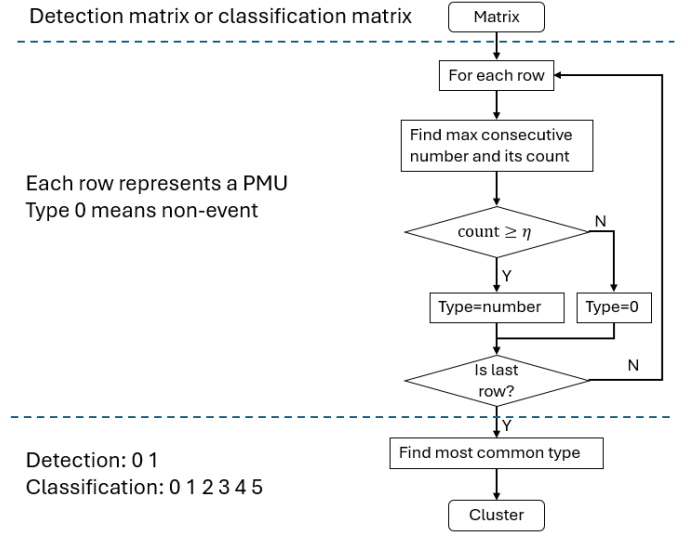

Fig. 6. Flowchart for event detection and classification identification.

#### 2) Decision Fusion

The decision fusion process integrates the outputs of the detection module and the classification module to produce the final event decision. As shown in Table I, the detection output provides a binary indication of whether an event is present, while the classification module predicts an event type if the data segment deviates from normal operation. The fusion rules ensure that both known and unknown events are handled consistently.

TABLE I
Decision Fusion

| Detection | Classification | Final |
|-----------|----------------|-------|
| 0 | 0 | 0 |
| 1 | 0 | 5 |
| 0 | 1,2,3,4 | 1,2,3,4 |
| 1 | 1,2,3,4 | 1,2,3,4 |

When both detection and classification outputs are zero, the system concludes that no event is present (Final = 0). If detection indicates an event (Detection = 1) but classification fails to assign a known label (Classification = 0), the segment is labeled as an unknown event type (Final = 5). Conversely, if detection does not indicate an event but classification predicts a type among the known categories {1, 2, 3, 4}, the final decision adopts the classification result, reflecting that the segment aligns with a recognized disturbance signature. Finally, when both detection and classification agree on the presence of an event, the classification outcome {1, 2, 3, 4} is carried forward as the final result.

This fusion logic provides a balanced mechanism to incorporate unsupervised detection with supervised classification. It allows the framework to capture both unknown disturbances, which are flagged as new event types, and known disturbances, which are mapped to predefined categories.

## III. SIMULATION RESULTS

This section presents the simulation results evaluating the performance of the proposed event detection and classification framework using both real-world synchrophasor data and simulated data. The model is tested against a diverse set of power system disturbances and benchmarked against benchmark methods.

### A. Model Training

The VAE-GAN model was trained to capture the normal operating behavior of power system dynamics using historical phasor measurement unit (PMU) data. Training was conducted in an unsupervised manner, where only normal (non-event) time-series segments were used to establish the latent distribution and reconstruction capabilities of the model. This approach enables the model to identify deviations associated with abnormal conditions without relying on labeled event data. Each PMU signal was first segmented into overlapping windows of fixed length (e.g., 5 seconds) with a defined stride (e.g., 1 second). These windowed segments served as input samples for training. The model was implemented using PyTorch and trained using the Adam optimizer with a learning rate of $2 \times 10^{-4}$. The batch size was set to 128, and early stopping was employed based on reconstruction loss on a validation set. To improve generalization, batch normalization was applied during training. Upon convergence, the model demonstrated the ability to reconstruct normal PMU patterns with low error while producing higher reconstruction and adversarial scores on eventful or abnormal sequences. This discrepancy forms the basis for effective anomaly detection in downstream applications.

### B. Benchmark Selection Clarification

To evaluate the effectiveness of the proposed framework, appropriate benchmarks were selected for both event detection and event classification tasks. For event detection, the proposed method is compared against established techniques, with emphasis on the Max Envelope method, a widely used baseline in transient detection due to its simplicity and effectiveness. This comparison highlights the benefits of integrating generative modeling and decision fusion over traditional threshold- and envelope-based approaches. For event classification, the goal is to validate the capability to handle unknown event types, a problem rarely addressed in prior work. Since conventional classifiers are designed for closed-set scenarios, direct comparison is not meaningful.

Instead, a controlled experiment is conducted by evaluating performance with and without the inclusion of unknown events, thereby demonstrating the unique contribution of the proposed framework in systematically categorizing previously unseen disturbances.

### C. Generative AI-Based Event Detection Result

#### 1) Event Detection Speed

High speed event detection is the key when dealing with massive amount of synchrophasor data. The scalability of the algorithm is tested on a laptop PC. It only takes around 5 seconds for the proposed detection to scan through a 1-minute synchrophasor dataset and create a list of events. It shows that the proposed detection method is able to detect events in a real-time manner.

#### 2) Event Detection on 5-second Window for Single PMU

To validate the effectiveness of the proposed VAE-GAN-based event detection framework, a series of detection experiments were conducted using labeled phasor measurement unit (PMU) time-series data containing both normal and abnormal system conditions.

Figure 7 illustrates the distribution of individual PMU data segments in the 2D feature space defined by the reconstruction error and discriminator error. As shown in the figures, event-related segments tend to occupy regions with higher values of either or both error metrics. In contrast, non-event samples cluster tightly in the lower-left quadrant of the error space, indicating high-fidelity reconstruction and low discriminator rejection.

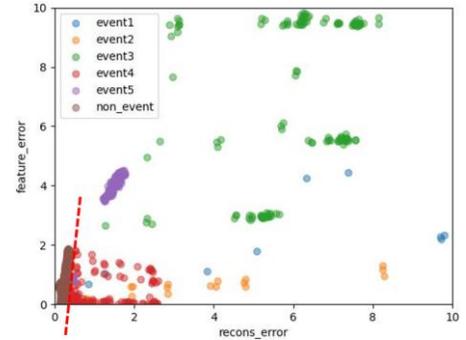

(a) Decision making based on linear threshold.

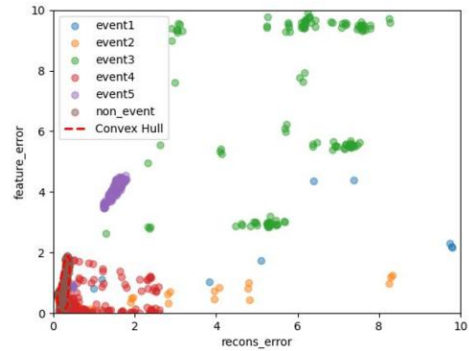

(b) Decision making based on Convex Hull.

Fig. 7. Event detection based on different decision-making methods.

In Figure 7 (a), a conceptual linear decision boundary is drawn to separate these two distributions, illustrating the discriminative capability of the learned scoring space. This visualization confirms that the joint error space provides a meaningful separation between normal and abnormal behavior, forming the basis for effective threshold-based detection. To determine an optimal detection threshold, a range of values was evaluated for their impact on binary classification accuracy. The accuracy increases rapidly as the threshold increases from 0 to approximately 0.025, reaching a peak of 97.30%. Beyond this point, accuracy gradually declines due to the increase in false negatives (i.e., undetected events). In Figure 7 (b), a Convex Hull is used to separate normal and abnormal scenarios, achieving a higher detection accuracy of 99.30% on 5-second measurements.

These results demonstrate that the proposed VAE-GAN model, when paired with an appropriate decision-making method, can effectively distinguish between event and non-event PMU behavior using only unsupervised training on normal data. The joint use of reconstruction and adversarial features enhances the model's sensitivity to subtle system changes while maintaining a low false-positive rate.

TABLE II
Event Detection Performance Comparison on 5-second Data

| Methods | Accuracy ($\frac{TP+TN}{Total}$) |
|---|---|
| Support Vector Machines | 86.24% |
| Decision Trees | 87.19% |
| k-Nearest Neighbors | 85.61% |
| Multilayer Perceptron | 90.48% |
| Convolutional Neural Network | 94.52% |
| Recurrent Neural Networks | 95.93% |
| VAE-GAN with linear threshold | 97.30% |
| VAE-GAN with Convex Hull | 99.30% |

Table II presents a comparison of event detection performance across several benchmark methods using 5-second PMU data segments. This comparison highlights not only the superiority of the proposed framework over established benchmarks but also the critical role of robust decision-making strategies in achieving state-of-the-art event detection performance.

### 3) Event Detection on 1-minute Window for Multiple PMUs

Figure 8 demonstrates the ability of the proposed VAE-GAN framework to detect and localize a power system event using 1-minute phasor measurement unit (PMU) data. The left subplot shows normalized voltage magnitude signals from multiple PMU channels over a 60-second interval. While most signals remain steady, a sharp drop occurs around timestamp 01:09:36, clearly indicating the presence of an event. A smaller disturbance is also visible just before the main deviation, suggesting early anomalous behavior.

The right subplot in Figure 8 presents a spatiotemporal anomaly matrix generated by the detection framework. Each row corresponds to a specific PMU channel, and each column represents a time step derived from overlapping sliding windows. Black squares mark the segments where the joint event score, based on reconstruction and adversarial errors, exceeds a predefined threshold. The resulting matrix highlights a well-localized cluster of detections aligned with the event timing and the affected channels. The framework successfully identifies not only the core event but also the precursor variations, offering both spatial and temporal precision in event characterization. This confirms the effectiveness of the proposed method in delivering interpretable, data-driven insights for real-time power system monitoring.

As a benchmark, the Max Envelope method [21] identifies disturbances by calculating the area between moving minimum and maximum bounds of power system signals. A larger envelope area typically indicates abnormal behavior.

To illustrate the practical differences between the two methods, several representative figures are included. Figures 9 and 10 illustrate qualitative differences between the two methods. Figure 9 showcases events detected exclusively by one method: (a) depicts events identified solely by the Max Envelope method, typically characterized by short, spike-like changes in single channels, while (b) highlights events detected only by VAE-GAN, which tend to involve subtle, multichannel disturbances that evolve over a longer duration. These differences underscore the limitations of envelope-based techniques when encountering spatially distributed or low-amplitude anomalies.

Figure 10 shows examples of events detected by both methods but with varying durations. In these cases, the VAE-GAN model generally captures a broader temporal scope of the disturbance, while the Max Envelope method often restricts its detection to the peak or central portion. This difference demonstrates VAE-GAN's ability to recognize prolonged or slowly progressing events, which is crucial for accurately assessing power system stability in real time.

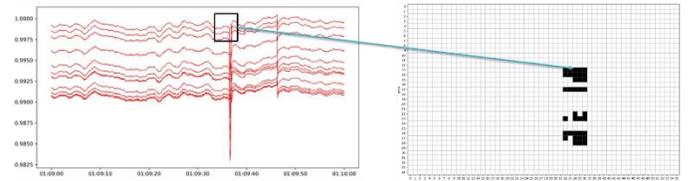

Fig. 8. Illustration of binary anomaly matrix.

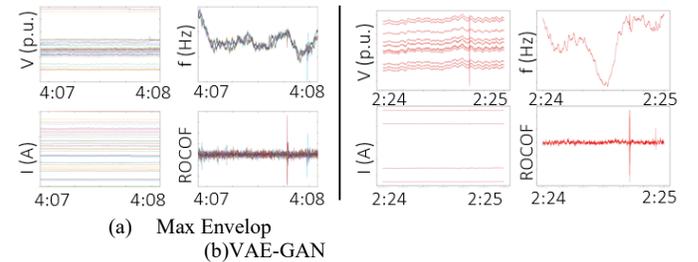

(a) Max Envelop          (b)VAE-GAN

Fig. 9. Example events detected by only one method.

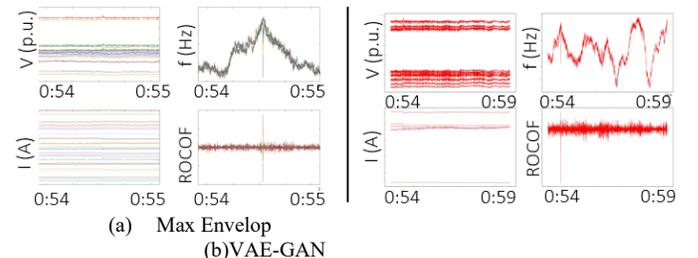

(a) Max Envelop          (b)VAE-GAN

Fig. 10. Example event detected by both methods with different durations.

Table III summarizes a comparative evaluation between the proposed VAE-GAN-based event detection approach and the Max Envelope method across multiple criteria. While the Max

Envelope technique is effective for identifying spike-like events and single-channel patterns, it struggles with more complex event dynamics. Specifically, it often misses long-term, low-magnitude, or multi-channel patterns due to its dependence on fixed time windows and threshold heuristics.

TABLE III
Performance Comparison

| Criteria | Max Envelope | VAE-GAN |
|---|---|---|
| Spike-like events | √ | √ |
| Long-term events | | √ |
| Pattern shown in one channel | √ | |
| Simplicity and interpretability | √ | |
| Subtle events | | √ |
| Generalization | | √ |

In contrast, the VAE-GAN framework demonstrates broader capabilities. It successfully captures not only sharp disturbances but also gradual, spatially distributed events across multiple PMUs. The model's unsupervised architecture and ability to learn directly from raw PMU waveforms contribute to its generalization power and sensitivity to subtle anomalies. Although the Max Envelope method remains appealing for its simplicity and interpretability, its effectiveness is limited to more diverse and data-rich operational scenarios.

### D. Event Classification Results

#### 1) 2-D Matrices Generation

Figure 11 presents an example of the detection and classification outcomes for a disturbance event. The left panel shows the voltage magnitude measurement, where a sharp deviation indicates the presence of an event. On the right, the detection matrix (top) provides a binary representation in which black squares correspond to abnormal segments flagged by the detection pathway. These blocks appear prominently in the middle region, indicating the time window where the event occurs, while the surrounding white regions represent normal operation. The classification matrix (bottom) extends this representation by assigning specific event categories to the abnormal segments, with non-zero values (1–4) mapped to different colors. At the edges of the event window, some variation in assigned event types appears due to the local differences captured by the sliding window, but in the middle of the event the majority of overlapping windows consistently identify the same type. This stability highlights the interpretability provided by the classification pathway.

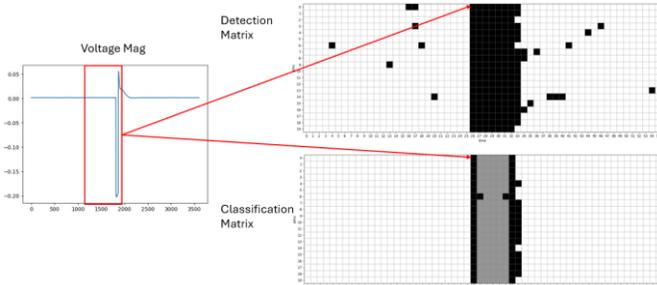

Fig. 11. 2D Matrix Example.

Taken together, the two matrices demonstrate complementary roles: the detection pathway ensures sensitivity to anomalies by highlighting the abnormal time segments, while the classification pathway provides additional detail by categorizing those segments into different event types.

#### 2) Event Identification and Decision Fusion

The binary detection matrix and non-binary classification matrix introduced earlier provide the foundational inputs for the next stage of event decision-making. While these matrices clearly capture when anomalies occur and how they may be categorized, further processing is required to transform them into reliable system-level event decisions. This is accomplished through the identification process, illustrated in Figure 6. This fusion step not only strengthens the robustness of known event classification but also enables the model to systematically handle unknown event types by assigning them to a new category.

#### 3) Event Classification for Unknown Event Types

Table IV demonstrates the effectiveness of the proposed method in improving event classification, particularly in the presence of unknown event types. Without the sliding window (top row), the classification accuracy is 81.89%. Accuracy means the ratio of correctly classified samples and total number of samples. As highlighted in the red box, samples belonging to the unseen event type 5 are scattered across multiple known categories. This behavior is expected from a purely supervised classifier, which lacks the ability to recognize unseen patterns and therefore reacts randomly when presented with unfamiliar inputs. While the detection module alone achieves 95.05% accuracy in flagging abnormal conditions, it cannot by itself distinguish between different event categories. The combined result modestly improves to 83.31%, but the challenge of unknown event misclassification remains evident.

With the sliding window mechanism (bottom row), both classification and detection accuracy improve. Classification accuracy rises slightly to 82.94%, while detection accuracy increases to 96.66%. More importantly, when combined through the decision fusion mechanism, overall accuracy jumps to 93.98%. Here, the sliding window enforces temporal consistency and reduces noise-driven misclassifications, where the decision fusion rules systematically integrate detection and classification results. This integration eliminates the randomness of supervised classifiers in dealing with unknown events, allowing the framework to correctly group these samples into a dedicated new category (type 5), rather than forcing them into existing classes.

These results highlight a key contribution of the proposed framework: the ability to classify previously unseen event types into a meaningful new category. By combining unsupervised detection with supervised classification under the sliding window and decision fusion mechanisms, the method provides both robustness to known events and adaptability to novel disturbances. This capability directly addresses one of the most important limitations of existing event classification methods in power systems, namely their inability to handle previously unseen event types.

TABLE IV
Event Classification Performance Improvement

| | **Classification Results** | **Detection Results** | **Combined Results** |
|---|---|---|---|
| **Without Proposed Method** |  Accuracy = 81.89% |  Accuracy = 95.08% |  Accuracy = 83.31% |
| **With Proposed Method** |  Accuracy = 82.94% |  Accuracy = 96.66% |  Accuracy = 93.98% |

Table V summarizes the per-class precision, recall, and F1-scores for event classification, comparing the baseline framework without the proposed method and the enhanced framework with it. For known events (Classes 0–4), both approaches achieve consistently high performance, with precision and recall values close to or above 0.9 in most cases. The most significant improvement, however, occurs for unknown events (Class 5). Without the proposed method, the classifier exhibits very low recall (0.106) and F1 (0.188), indicating that most unknown events are misclassified into existing categories. By contrast, with the proposed method, precision for Class 5 increases to 0.971, recall rises to 0.667, and F1 improves substantially to 0.791. This demonstrates that the integration of the sliding window and decision fusion mechanisms effectively mitigates the randomness of supervised classification when dealing with unseen events, enabling the framework to assign them systematically into a new category while preserving strong performance for known event types.

TABLE V
Per-Class Evaluation Before and After Proposed Method

| | Without Proposed Method | | | With Proposed Method | | |
|---|---|---|---|---|---|---|
| Class | Precision | Recall | F1 | Precision | Recall | F1 |
| 0 | 0.781 | 0.980 | 0.869 | 0.833 | 1.000 | 0.909 |
| 1 | 0.882 | 0.940 | 0.910 | 1.000 | 0.980 | 0.990 |
| 2 | 0.645 | 0.974 | 0.776 | 0.877 | 1.000 | 0.934 |
| 3 | 1.000 | 0.995 | 0.998 | 1.000 | 0.990 | 0.995 |
| 4 | 0.939 | 1.000 | 0.968 | 1.000 | 1.000 | 1.000 |
| 5 | 0.840 | 0.106 | 0.188 | 0.971 | 0.667 | 0.791 |

*4) Ablation Study*

To better understand the contribution of the decision fusion and sliding window mechanisms in the proposed framework, an ablation study was performed by evaluating four different scenarios, as shown in Table VI.

TABLE VI
Event Classification Performance in Different Scenarios

| | Decision Fusion | Sliding Window | Detection Accuracy | Classification Accuracy |
|---|---|---|---|---|
| Scenario 1 | - | - | - | 81.89% |
| Scenario 2 | √ | - | 95.08% | 83.31% |
| Scenario 3 | - | √ | - | 82.94% |
| Scenario 4 | √ | √ | 96.66% | 93.98% |

Scenario 1 represents the baseline configuration without either component, yielding a classification accuracy of 81.89%. This baseline reflects the limitations of a purely supervised classifier: while it performs reasonably well for known events, it struggles to handle unseen events, leading to reduced overall accuracy.

In Scenario 2, decision fusion is introduced without the sliding window. This significantly improves detection accuracy to 95.08% and raises classification accuracy to 83.31%. The gain is attributed to the fusion mechanism, which integrates binary anomaly detection with supervised classification, thereby reducing the misclassification of unknown events into existing categories. However, the absence of temporal smoothing means that some misclassifications still occur due to noise and local variations in short data segments.

Scenario 3 incorporates the sliding window without decision fusion. Here, classification accuracy improves modestly to 82.94%. The sliding window enforces temporal consistency by aggregating decisions across overlapping segments, which reduces noise-driven errors at the boundaries of events. While this enhances robustness, the lack of decision fusion means the classifier still cannot systematically handle unknown events, limiting the overall improvement.

Scenario 4 combines both decision fusion and sliding window mechanisms, achieving the best performance with a detection accuracy of 96.66% and a classification accuracy of 93.98%. In this configuration, the sliding window ensures temporal stability, while decision fusion integrates detection and classification to correctly recognize both known and unknown events. The synergy between these two mechanisms eliminates the randomness of purely supervised approaches when faced with unseen patterns, leading to consistent improvements across all event types.

Overall, the results demonstrate that while each mechanism individually contributes to improved classification, their combination provides the most substantial benefit. This confirms the necessity of jointly leveraging temporal processing and multi-path decision-making to achieve robust and generalizable event classification in synchrophasor data.

## IV. CONCLUSION

This paper proposed a generative modeling framework for event detection and classification in power systems using synchrophasor data. A VAE–GAN model was employed to learn the distribution of normal operating conditions, with reconstruction and discriminator errors serving as anomaly indicators. Two decision strategies, threshold-based and convex hull–based, were developed, achieving detection accuracies of 97.3% and 99.3%, respectively, which clearly surpass conventional baselines. These results demonstrate the advantage of combining generative modeling with robust geometric decision boundaries.

For event classification, the proposed framework addresses a fundamental limitation of conventional supervised approaches: the inability to handle previously unseen event types. The integration of sliding-window temporal processing and decision fusion improved classification accuracy to 93.9%, confirming that the framework can systematically assign unseen disturbances to a dedicated new category rather than misclassifying them into known types.

Overall, the results validate that the proposed method not only achieves state-of-the-art detection accuracy but also provides a novel capability for unknown event classification, a feature largely absent in prior work. Future research will extend this framework to larger PMU networks and explore online adaptation for evolving grid conditions.